\documentclass[runningheads]{llncs}

\usepackage{silence}
\usepackage[font=small, labelfont=bf]{caption}

\usepackage[T1]{fontenc}
\usepackage[utf8]{inputenc}
\usepackage{acro}
\usepackage{adjustbox}
\usepackage{booktabs}
\usepackage{comment}
\usepackage{todonotes}
\usepackage{colortbl}
\usepackage{hyperref}

\newcolumntype{a}{>{\columncolor{gray!10!white}}c}
\newcolumntype{x}{>{\columncolor{green!10!white}}c}
\newcolumntype{y}{>{\columncolor{blue!10!white}}c}
\newcolumntype{z}{>{\columncolor{yellow!10!white}}c}
\newcolumntype{v}{>{\columncolor{red!10!white}}c}
\definecolor{OliveGreen}{rgb}{0,0.6,0}
\definecolor{ForestGreen}{RGB}{34,139,34}
\definecolor{myblue}{RGB}{37,165,203}
\definecolor{FAUblue}{rgb}{0.000, 0.2196, 0.3961}
\definecolor{myred}{RGB}{175,32,67}

\colorlet{backgroundcol}{cyan!10!white}

\usepackage{graphicx}
\usepackage{enumitem}
\usepackage{color}
\usepackage{xcolor}
\usepackage{listings}
\definecolor{codegreen}{rgb}{0,0.6,0}
\definecolor{codegray}{rgb}{0.5,0.5,0.5}
\definecolor{codepurple}{rgb}{0.58,0.5,0.82}
\definecolor{backcolour}{rgb}{0.95,0.95,0.92}

\lstdefinestyle{mystyle}{
    backgroundcolor=\color{white},   
    commentstyle=\color{codegreen},
    keywordstyle=\color{magenta}\bfseries,
    moredelim=[is][\color{magenta}\bfseries]{@}{@},
    numberstyle=\tiny\color{codegray},
    stringstyle=\color{codepurple},
    basicstyle=\C\footnotesize,
    frame           = tb,         
    framerule       = 0.6pt,      
    rulecolor       = \color{black},
    framesep        = 0.4em,      
    xleftmargin     = 2em,
    framexleftmargin= 2em,
    breakatwhitespace=false,         
    breaklines=true,                 
    captionpos=b,                    
    keepspaces=true,                 
    numbers=left,                    
    numbersep=5pt,                  
    showspaces=false,                
    showstringspaces=false,
    showtabs=false,                  
    tabsize=2
}

\usepackage{amsmath,amssymb,amsfonts}

\usepackage{wrapfig}
\usepackage{graphicx}
\usepackage{soul}

\usepackage{multirow}

\begin{document}

\title{High-Performance Resilient Multi-GPU Hybrid Particle-in-Cell Monte Carlo Simulations at Scale}

%
\titlerunning{High-Performance Resilient Multi-GPU Hybrid PIC MC Simulations at Scale}
%

\author{Jeremy J. Williams\inst{1}\and
Stefan Costea\inst{2} \and
David Tskhakaya \inst{3} \and
Leon Kos \inst{2} \and
Ales Podolnik \inst{3} \and
Jakub Hromadka\inst{3} \and
Jordy Trilaksono \inst{4} \and
Yi Ju \inst{7} \and
Kallia Chronaki \inst{5} \and
Evangelos Gkolantas \inst{5} \and
Vassilis Papaefstathiou \inst{5} \and
Allen D. Malony \inst{6} \and
Sameer Shende \inst{6} \and
Frank Jenko \inst{2} \and
Erwin Laure \inst{7} \and
Stefano Markidis\inst{1} }
\authorrunning{Jeremy J. Williams et al.}
%
\institute{KTH Royal Institute of Technology, Stockholm, Sweden \and
Faculty of Mechanical Engineering, University of Ljubljana, Ljubljana, Slovenia \and
Institute of Plasma Physics of the CAS, Prague, Czech Republic  \and
Max Planck Institute for Plasma Physics, Garching, Germany \and
Foundation for Research and Technology Hellas, Heraklion, Greece \and 
University of Oregon, Eugene, Oregon, The United States of America \and
Max Planck Computing and Data Facility, Garching, Germany}
\maketitle              
\begin{abstract}
The increasing demand for high-performance computing in plasma physics has driven scalable and resilient simulation methods capable of efficiently exploiting modern multi-GPU architectures. This work extends a portable hybrid MPI+OpenMP implementation of BIT1, focusing on high-performance resilience for accelerated Particle-in-Cell (PIC) Monte Carlo (MC) simulations under both uniform and non-uniform load conditions. Scalable particle load balancing and robust checkpoint/restart mechanisms across Nvidia and AMD accelerators are integrated with standardized I/O using openPMD and ADIOS2. This leverages BP4 for high-performance file-based checkpointing and SST for in-memory data streaming, enabling efficient data movement, resilient large-scale execution, seamless continuation from existing checkpoints, and effective handling of computational and I/O workloads. Advanced HPC profiling and tracing tools, including Nvidia Nsight Systems and AMD ROC-Profiler with Perfetto, provide detailed insights into computation, communication, and system-level behavior for optimization. Performance results on Frontier (OLCF-5), MN5, and LUMI-G demonstrate strong and weak scaling up to 800 GPUs, validating the framework for large-scale PIC MC simulations, while in-situ analysis and visualization using scalable I/O further enhance scientific insight without interrupting multi-GPU execution on current and future exascale systems.

\keywords{Heterogeneous Computing \and Hybrid MPI+OpenMP \and  BIT1 \and Nvidia \and AMD \and HPC Resilience \and Checkpoint/Restart \and Particle Load Balancing \and Asynchronous Execution \and Large-Scale PIC MC Simulations}
\end{abstract}


\section{Introduction}
Plasma physics applications are increasingly pushing the limits of modern high-performance computing systems, where scalability and resilience on heterogeneous architectures are essential for the next-generation systems~\cite{kurzyniec2005failure}. Particle-in-Cell (PIC) Monte Carlo (MC) simulations are a key computational method in this field. However, their performance is strongly affected by dynamic particle distributions, irregular memory access patterns, and the need for efficient data movement and resilience~\cite{iakymchuk2016particle}. 

In this work, we extend a portable hybrid MPI+OpenMP implementation of BIT1 with scalable particle load balancing and checkpoint/restart mechanisms for high-performance resilience in large-scale multi-GPU PIC MC simulations across Nvidia and AMD accelerator architectures. These capabilities are unified through a standardized I/O layer based on openPMD and ADIOS2. This uses BP4 for high-throughput file-based checkpointing and SST for in-memory streaming, enabling efficient data movement, resilient large-scale, long-running execution, and effective management of computational and I/O workloads.

The contributions of this work are the following:
\begin{itemize}[leftmargin=*]
\item  We design and implement a portable, resilient hybrid MPI+OpenMP extension of BIT1 for improved performance on modern HPC systems under both strong and weak scaling, supporting uniform and non-uniform load conditions.
\item We introduce scalable particle load balancing and checkpoint/restart mechanisms for resilient large-scale execution, seamless continuation from checkpoints, and efficient management of computational and I/O workloads across Nvidia and AMD GPU architectures.
\item We integrate a standardized I/O layer using openPMD and ADIOS2, leveraging BP4 for high-throughput checkpointing and SST for in-memory streaming, enabling efficient data movement, continuous execution, and consistent state management in large-scale multi-GPU executions.
\item We utilize a customized Python script with the openPMD API and ADIOS2 BP4 and SST backends for real-time checkpoint analysis and visualization of hybrid BIT1 file-based I/O (from disk) and in-memory data streaming, tailored for hybrid BIT1 output, without interrupting the simulation at scale. 
\end{itemize}


\section{Background}
The PIC method is a numerical technique used to model plasma behavior by simulating particle dynamics in one to three spatial dimensions (1–3D) while typically resolving three-dimensional velocity space (3V). For plasma edge applications, PIC is often coupled with MC routines to model particle collisions and interactions with device walls. The PIC cycle normally begins with initialization of the simulation domain, particle load, and physical parameters, followed by repeated timesteps consisting of: (i) particle mover, updating particle positions and velocities; (ii) deposition to the grid, mapping particle quantities to compute charge and current densities; (iii) field solver, computing electromagnetic or electrostatic fields; and (iv) particle force evaluation, interpolating fields back to particles, while additional steps such as density smoothing and MC collision handling improve numerical stability and capture collisional effects. The cycle is repeated over many timesteps to simulate plasma evolution, with each timestep representing a small fraction of the characteristic plasma timescale to ensure accurate resolution of the underlying dynamics, as detailed in~\cite{tskhakaya2010pic,williams2023leveraging,williams2025integrating,williams2026multigpu}.

In this work, we use the Berkeley Innsbruck Tbilisi 1D3V (BIT1) code for large-scale plasma simulations on HPC systems, targeting plasma edge and tokamak divertor physics~\cite{tskhakaya2010pic}. BIT1 is a 1D3V electrostatic PIC MC code derived from Birdsall’s \texttt{XPDP1} code~\cite{verboncoeur1993simultaneous}, written in C ($\approx$31,600 lines of code) with native Poisson solver, particle mover, and smoothing operators. BIT1 employs the “natural sorting” method~\cite{tskhakaya2007optimization}, which significantly accelerates collision operators but increases memory usage and complicates GPU offloading and load balancing in high-density regions~\cite{williams2023leveraging,williams2025accelerating}. To improve portability and scalability on heterogeneous HPC systems, BIT1 is evolving toward a hybrid MPI + OpenMP model that reduces communication overhead and improves performance on multi-core and multi-GPU architectures~\cite{williams2024optimizing,williams2025accelerating,williams2026multigpu}.

\subsection{openPMD Standard, openPMD-api, \& ADIOS Version 2}
The openPMD standard~\cite{openPMDstandard} provides a portable format for particle and mesh-based simulation data, enabling interoperability across scientific codes and supporting backends such as HDF5, ADIOS1, ADIOS2, and JSON. The openPMD-api~\cite{openPMDapi} supports serial and MPI-parallel I/O by organizing data into records (particles or fields) over time iterations, forming complete simulation series. ADIOS2 (Adaptable Input Output System v2) is a high-performance I/O framework for checkpoint/restart, in-situ analysis, and in-memory streaming across C, C++, Fortran, and Python~\cite{williams2025integrating}, offering multiple engines including BP5, BP4, BP3, HDF5, SST, SSC, DataMan, DataSpaces, and Inline. 


\section{Related Work} 
The transition to exascale computing has accelerated the adoption of heterogeneous architectures and hybrid programming models for large-scale particle simulations. Prior work has demonstrated efficient hybrid coupling of CPU and GPU components, emphasizing communication hiding, CUDA-aware MPI, and partitioning strategies to minimize overhead and achieve strong scaling on thousands of GPUs~\cite{kemmler2025efficiency}, highlighting the importance of efficient data movement and balanced resource utilization. Load balancing remains a critical challenge due to dynamic particle migration and non-uniform distributions, with recent DSMC/PIC approaches employing adaptive grid partitioning and runtime particle redistribution to maintain scalability across thousands of processes~\cite{qiu2022parallelizing}. Similarly, exascale PIC frameworks such as WarpX demonstrate multi-level parallelization and efficient load balancing, enabling large-scale plasma simulations with improved performance~\cite{fedeli2022pushing}. Resilience is increasingly important as system failure rates grow, while checkpoint/restart is widely used, it introduces overhead at scale. Recent work addresses this through GPU kernel-level checkpointing for preemptive recovery~\cite{eiling2023checkpoint} and by combining checkpointing with MPI process replication to improve fault tolerance efficiency~\cite{joshi2024combining}, highlighting the need for scalable, low-overhead resilience in heterogeneous multi-GPU simulations.


\section{Methodology \& Experimental Setup}
In this work, we focus on enabling high-performance resilience for large-scale multi-GPU PIC MC simulations by introducing hybrid BIT1 enhancements, including scalable load balancing, checkpoint/restart, and integrated I/O strategies for efficient execution at scale.

\subsection{High-Performance Resilient Multi-GPU Hybrid BIT1}
To improve scalability, resilience, and execution efficiency at scale, key improvements were integrated using openPMD and ADIOS2 into hybrid BIT1 for large-scale multi-GPU PIC MC simulations.

\noindent\textbf{Particle Load Balancing (PLB).}
\noindent As presented by Williams et al.~\cite{williams2023leveraging,williams2024enabling,williams2025integrating}, hybrid BIT1 initially relied on serial I/O and was later extended with the openPMD-api to support parallel I/O using BP4 (high-throughput file-based checkpointing) and SST (in-memory streaming). This improved I/O functionality by consolidating write operations in \texttt{write\_parallel.cpp}. Subsequent studies~\cite{williams2025integrating,williams2026multigpu} identified load imbalance, addressed through PLB and restoration using openPMD in \texttt{read\_parallel.cpp}, enabling recovery of particle and mesh data and improved MPI workload distribution.
PLB is controlled in \texttt{bit1.cpp} using the \texttt{load\_balance} flag (\texttt{1} enabled, \texttt{0} disabled by default). Without PLB, each MPI rank processes a fixed interval $[rmpi*nc:(rmpi+1)*nc]$, where \texttt{rmpi} is the rank and \texttt{nc} is the number of cells per rank, often leading to uneven particle distributions. With PLB enabled, particles are redistributed across ranks, producing variable cell intervals computed by \texttt{new\_nc\_openPMD()} in \texttt{start.cpp} before parameterization and memory allocation. MPI rank 0 determines particle distribution, cell intervals, and data offsets, and distributes this information to all ranks, which update their local parameters accordingly. Each rank then loads mesh and particle data using \texttt{restore\_balanced\_openPMD()}, after which the openPMD iteration is closed to trigger \texttt{flush()}, and simulation arrays are updated with the restored data, improving workload balance and overall execution efficiency.

\noindent\textbf{Checkpoint/Restart (C/R).}
\noindent At the checkpoint of hybrid BIT1 simulation runs, particle meshes are stored individually for each particle species and are read following the procedure improved by Williams et al.~\cite{williams2024enabling,williams2025integrating,williams2026multigpu}. Each MPI rank opens the openPMD series object with file path, access mode, and communicator. Iteration 0 captures the latest simulation state for the required restart purposes. Particle counts per cell are retrieved and stored in a vector, followed by a flush for I/O efficiency. These counts are then used to compute local extents, offsets, and data loading regions per rank, enabling reconstruction of particle positions and velocities. After data loading is completed, the iteration and series are closed, triggering a final flush. Particle counts, positions, and velocities are updated, and temporary vectors are cleared to optimize memory usage, ensuring particle meshes are correctly restored for hybrid BIT1 simulations.

%

\subsection{Experimental Setup \& Use Case}
In this work, we focus on evaluating a portable, resilient hybrid MPI+OpenMP extension of BIT1 for improved performance on modern HPC systems under uniform and non-uniform load conditions. We simulate hybrid BIT1 on four HPC systems (Table~\ref{tab:hpc_systems_columns}) using the key parameters in Table~\ref{tab:bit1_input_parameters}. 

\begin{table}[!ht]
\vspace{-0.4cm} 
\centering
\renewcommand{\arraystretch}{1.3}
\begin{adjustbox}{max width=\columnwidth,center}
\footnotesize
\begin{tabular}{cccc>{\columncolor{gray!20}}c}
\hline
\textbf{Characteristic} & 
\textbf{Dardel CPU/GPU} & 
\textbf{MareNostrum5 (MN5) GPP/ACC} & 
\textbf{LUMI-C / LUMI-G} & 
\textbf{Frontier (OLCF-5)} \\
\hline
HPC System  & HPE Cray EX Supercomputer & Pre-Exascale EuroHPC Supercomputer & Pre-Exascale EuroHPC Supercomputer &  HPE Cray EX Exascale Supercomputer \\
Processor  & 2 × AMD EPYC Zen2 (64 cores) & 2 × Intel Sapphire Rapids 8480+ (56 cores) & 2 × AMD EPYC 7763 (64 cores) &  1 × AMD EPYC (64 cores) \\
CPU Nodes  & 1,278 & 6,480 & 2,048 &  -- \\
CPU + GPU Nodes & 62 & 1,120 & 2,978 & 9,856 \\
GPUs per Node & 4 x AMD MI250X (w/ 2 x GCDs)  & 4 × Nvidia H100 & 4 x AMD MI250X (w/ 2 x GCDs) & 4 x AMD MI250X (w/ 2 x GCDs) \\
Memory per Node & 256~GB – 2~TB & 128~GB HBM – 1~TB & 256~GB – 1~TB & 512~GB DDR4 \\
Network & Slingshot 200~GB/s & NDR200 InfiniBand & Slingshot-11 & Slingshot up to 800~Gb/s \\
Storage & 679~PB  & 248~PB Online / 402~PB Archive & 117 PB  & 679~PB \\
Location  & KTH Royal Institute of Technology & Barcelona Supercomputing Center & CSC – IT Center for Science &  Oak Ridge Leadership Computing Facility \\
\hline
\end{tabular}
\end{adjustbox}
\vspace{1mm}  
\caption{Key characteristics of the HPC systems used in performance tests.}
\label{tab:hpc_systems_columns}
\vspace{-1.2cm} 
\end{table}

\begin{table}[!ht]
\vspace{-0.4cm} 
\centering
\renewcommand{\arraystretch}{1.3}
\begin{adjustbox}{max width=\columnwidth,center}
\tiny
\begin{tabular}{c
                p{0.65\columnwidth} 
                >{\columncolor{gray!20}\centering\arraybackslash}p{0.15\columnwidth} 
                >{\columncolor{gray!20}\centering\arraybackslash}p{0.15\columnwidth}} 
\hline
\textbf{Parameter} & \textbf{Description} & \textbf{Uniform (Particle) Load} & \textbf{Non-Uniform (Particle) Load} \\
\hline
\texttt{slow} & Sets the plasma profiles and distribution function diagnostics (default=0). & 0 & 0 \\
\texttt{datfile} & Enables time-averaged diagnostics of plasma and particle distributions. & 0/100 & 100 \\
\texttt{dmpstep} & Defines when the simulation state is written for restart or checkpointing. & 0/1000 & 1000 \\
\texttt{mvflag} & Specifies a diagnostic snapshot averaged over \texttt{mvflag} time steps. & 0/100 & 100 \\
\texttt{mvStep} & Sets the interval between successive diagnostic outputs. & 0/500 & 500 \\
\texttt{Last\_step} & Specifies the final time step at which the simulation terminates. & 200/2000 & 10000 \\
\texttt{origdmp} & Sets the format; 1=File (Serial) I/O \& 2=openPMD (Parallel) BP4/SST. & 1/2 & 1/2 \\
\hline
\end{tabular}
\end{adjustbox}
\vspace{1mm}
\caption{Main input parameters for simulation output and control, with key flags for uniform \& non-uniform load conditions.~\cite{williams2024enabling,williams2024understanding}}
\label{tab:bit1_input_parameters}
\vspace{-1.0cm} 
\end{table}

Our simulations target the formation of a high-density sheath in front of so-called \emph{divertor} plates in future magnetic confinement fusion devices, such as ITER and DEMO, using two simulation setups with the same physical model but different initial loading and plasma source configurations.

More precisely, the two simulation setups are as followed in Table~\ref{tab:sheath_cases}.

\begin{table}[!ht]
\vspace{-0.4cm} 
\centering
\renewcommand{\arraystretch}{1.5}
\begin{adjustbox}{max width=\columnwidth,center}
\tiny
\begin{tabular}{p{0.48\columnwidth}|
                p{0.48\columnwidth}}
\hline
{\centering \textbf{High-Density Sheath (Uniform Load)}\par} &
{\centering \textbf{High-Density Sheath (Non-Uniform Load)}\par} \\
\hline

A double\allowbreak-bounded, magnetized plasma layer between two walls is considered, initially filled with electrons and $D^+$ ions. Plasma is absorbed at the walls (depositing charge), recycling $D^+$ ions into $D$ neutrals and forming a sheath. The 1D geometry has 3M cells, three species, and an initial 200 particles per cell per charged species ($\approx$1.2B total). Simulations run for 100K time steps unless stated otherwise. Baseline simulation: five nodes (Dardel), 640 MPI processes, exceeding single-node memory. 
&
A double-bounded, magnetized plasma layer between two walls is considered. The domain is initially empty and a volumetric plasma source is placed in the left half of the domain, mimicking downstream plasma flow. Plasma fills the domain and is absorbed at the walls, depositing charge and forming a sheath. $D^+$ ions are recycled into $D$ neutrals upon wall interaction. The 1D geometry has 3M cells. Simulations run for 100K time steps unless stated otherwise. Baseline simulation: 10 nodes (Dardel), 1280 MPI processes, requiring up to 2× the memory needed for a uniform load simulation. \\

\hline
\end{tabular}
\end{adjustbox}
\vspace{1mm}
\caption{High-density sheath simulation cases (uniform and non-uniform load)~\cite{williams2023leveraging,williams2026multigpu}.}
\label{tab:sheath_cases}
\vspace{-1.3cm} 
\end{table}


\section{Performance Results, Analysis \& Visualization}
In this work, we evaluate the impact and performance of integrating openPMD with a portable, resilient hybrid BIT1 using the ADIOS2 BP4 and SST backends.

\subsection{Profiling \& Understanding Hybrid BIT1 up to 8 GPUs}
We begin by utilizing \texttt{Nvidia Nsight Systems} and \texttt{AMD ROC-Profiler with Perfetto} to gain insights into GPU and CPU execution, memory operations, and system-level performance bottlenecks to support application optimization. 

\begin{figure}[!ht]
    \vspace{0cm} 
    \begin{center}
        \includegraphics[width=\linewidth]{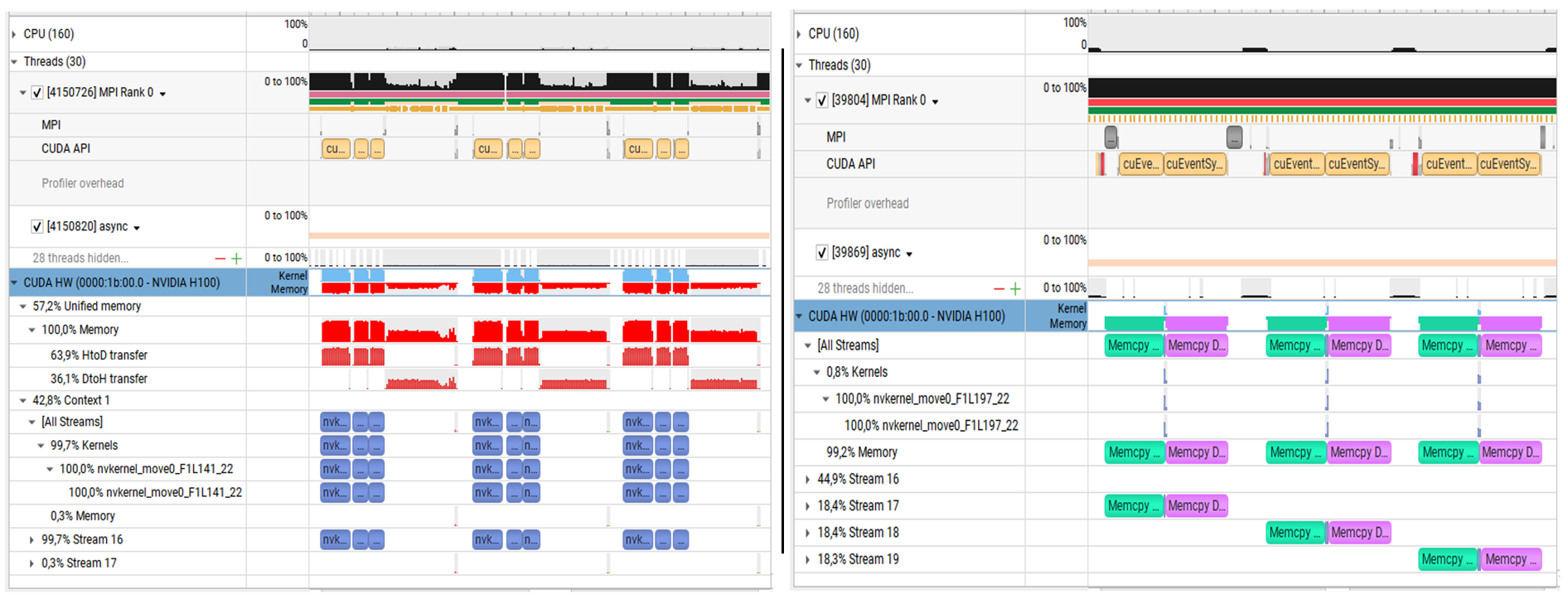}
        \caption{Nvidia Nsight Systems OpenMP UM (Left) and PinM (Right) View - GPU Porting of BIT1 Mover Function on MN5 ACC  for 3 timesteps.} 
        \label{NVIDIA_NSIGHT_SYSTEMS_OpenMP_PinM_Three_Time_Steps}
    \end{center}
    \vspace{-0.6cm} 
\end{figure}

Fig.~\ref{NVIDIA_NSIGHT_SYSTEMS_OpenMP_PinM_Three_Time_Steps} shows profiling results obtained with Nvidia Nsight Systems on MN5 ACC (Nvidia GPUs) during the development of the BIT1 GPU offloading strategy~\cite{williams2024optimizing,williams2025accelerating}. Unified Memory (UM) introduced frequent page faults, implicit data migrations, and serialization, leading to higher latency, fragmented GPU kernel execution, and longer runtimes, whereas Pinned (Page-Locked) Host Memory (PinM) eliminated page faults and enabled explicit, high-throughput host-to-device (HtoD) transfers, resulting in more compact and contiguous kernel execution, and lower data-movement overheads with improved performance. 

\begin{figure}[!ht]
    \vspace{-0.4cm} 
    \begin{center}
        \includegraphics[width=\linewidth]{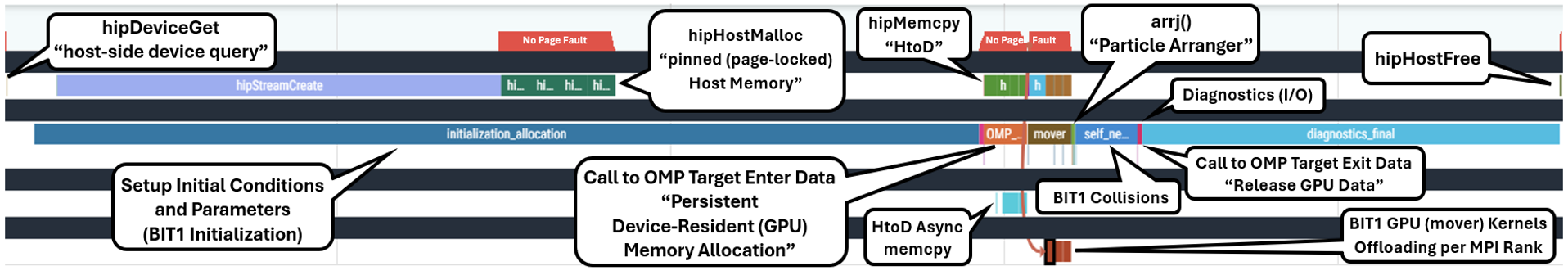}
        \caption{AMD ROC-Profiler (rocprof + Perfetto) OpenMP PinM (HSA and HIP activity) View - GPU Porting of BIT1 Mover Function on Dardel GPU (with corresponding confirmation traces on both LUMI-G and Frontier)  for 1 timestep.} 
        \label{AMD_Persident_GPU_One_Time_Step_One_GPU}
    \end{center}
    \vspace{-0.8cm} 
\end{figure}

Building on the UM to PinM transition, compile-time pinned memory used on MN5 ACC (Nvidia GPUs) proved difficult to port to AMD GPU architectures and introduced overhead by pinning all data structures rather than only large, performance-critical arrays. To address this, hybrid BIT1 adopts a portable approach~\cite{williams2026multigpu} on Dardel, LUMI, and Frontier, using predefined OpenMP memory allocators with pinned memory traits~\cite{neth2021beyond,sewall2016modern} for selective allocation in the particle mover (\texttt{move0, moveb}) and arranger (\texttt{arrj, narrj}) functions, improving data-transfer efficiency, and enabling GPU interoperability through direct device-pointer access using OpenMP \texttt{use\_device\_ptr} and \texttt{is\_device\_ptr} clauses, removing redundant mappings for improved asynchronous execution. Profiling with AMD ROC-Profiler (\texttt{rocprof}), as shown in Fig.~\ref{AMD_Persident_GPU_One_Time_Step_One_GPU}, captures detailed HSA and HIP activity, including ROCTX regions, asynchronous data transfers, and GPU kernel execution, with traces visualized in \texttt{Perfetto} for fine-grained analysis of overlap, kernel concurrency, and data movement. 

\subsection{Developing \& Porting Hybrid BIT1 up to 320 GPUs}
After profiling and extending BIT1 to a task-based hybrid version with a focus on mitigating load imbalance and optimizing resource utilization~\cite{williams2023leveraging,williams2025accelerating}, we ported it to MN5 ACC on 5 nodes (four GPUs per node) to analyze the impact of a contiguous 1D data layout, OpenMP target parallelism, and multi-GPU asynchronous execution.

\begin{figure}[!ht]
    \vspace{0cm} 
    \begin{center}
        \includegraphics[width=\textwidth]{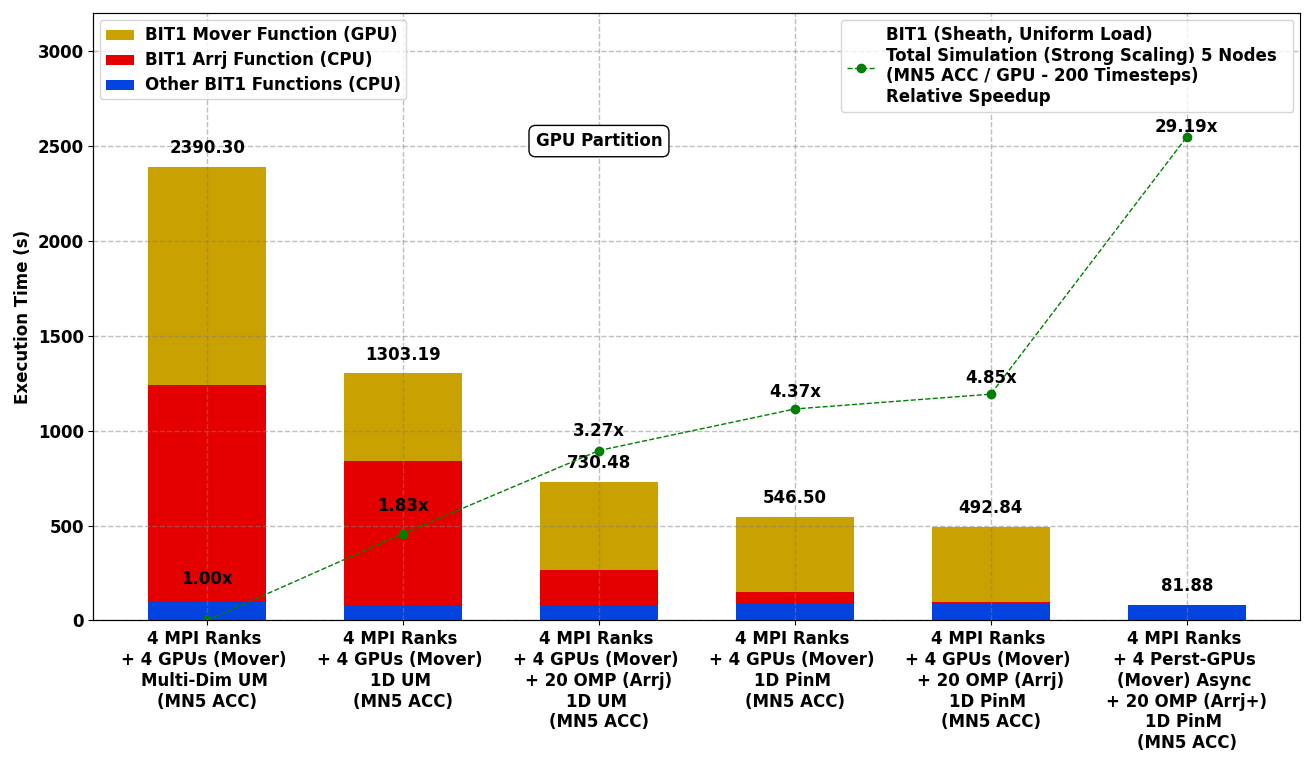}
     \end{center}
     \caption{Hybrid BIT1 (Sheath, Uniform Load) Total Simulation (Development Progression) Strong Scaling on 5 Nodes (20 GPUs) on MN5 ACC for 200 timesteps.} 
     \label{BIT1_Strong_Scaling_5_Node_MN5_Execution_Time_UM_and_PinM_GPU_Development_Work}
     \vspace{-0.4cm} 
\end{figure}

As seen in Fig.~\ref{BIT1_Strong_Scaling_5_Node_MN5_Execution_Time_UM_and_PinM_GPU_Development_Work}, the original BIT1 version, using 4 MPI + 4 GPUs (Mover) with a 3D data layout and UM, executes with a total simulation time of $\approx$2390~s, with the \texttt{mover} function dominating at $\approx$1150~s and the \texttt{arrj} function taking $\approx$1135~s. Transitioning to a 1D UM layout improves memory access, reducing the total time to $\approx$1303~s and lowering the \texttt{mover} and \texttt{arrj} times to $\approx$462~s and $\approx$767~s, respectively. Introducing OpenMP thread parallelism to the \texttt{arrj} function further decreases its execution time to $\approx$186~s, bringing the total simulation time to $\approx$731~s, while the \texttt{mover} executes for $\approx$467~s. Using PinM for the \texttt{mover} reduces data transfer overhead, decreasing the total time to $\approx$547~s, with \texttt{arrj} and \texttt{mover} at $\approx$59~s and $\approx$397~s. Combining PinM with OpenMP \texttt{arrj} parallelism further reduces the total time to $\approx$493~s, with \texttt{arrj} reduced to $\approx$8~s. Finally, the fully optimized version with persistent GPU memory, asynchronous \texttt{mover} execution, and OpenMP \texttt{arrj} parallelism reduces the \texttt{mover} time to near zero ($\approx$0.07~s) and \texttt{arrj} to $\approx$2~s, resulting in a total simulation time of around $\approx$82~s, corresponding to a 29.19× speedup over the original BIT1 (3D data layout + UM) simulation.

Next, we evaluate the portable, multi-GPU hybrid MPI\allowbreak+OpenMP asynchronous version of BIT1 under non-uniform load conditions. This is performed using a 10K time step sheath simulation on 40 nodes (up to 320 GPUs), with the parameters listed in Table~\ref{tab:bit1_input_parameters}, representing a heavier diagnostic workload with frequent time dependent output and checkpointing, which heavily stresses computation, communication, I/O, in-situ analysis and visualization, and most importantly, the file system.

\begin{figure}[!ht]
    \vspace{-0.4cm} 
    \begin{center}
        \includegraphics[width=\linewidth]{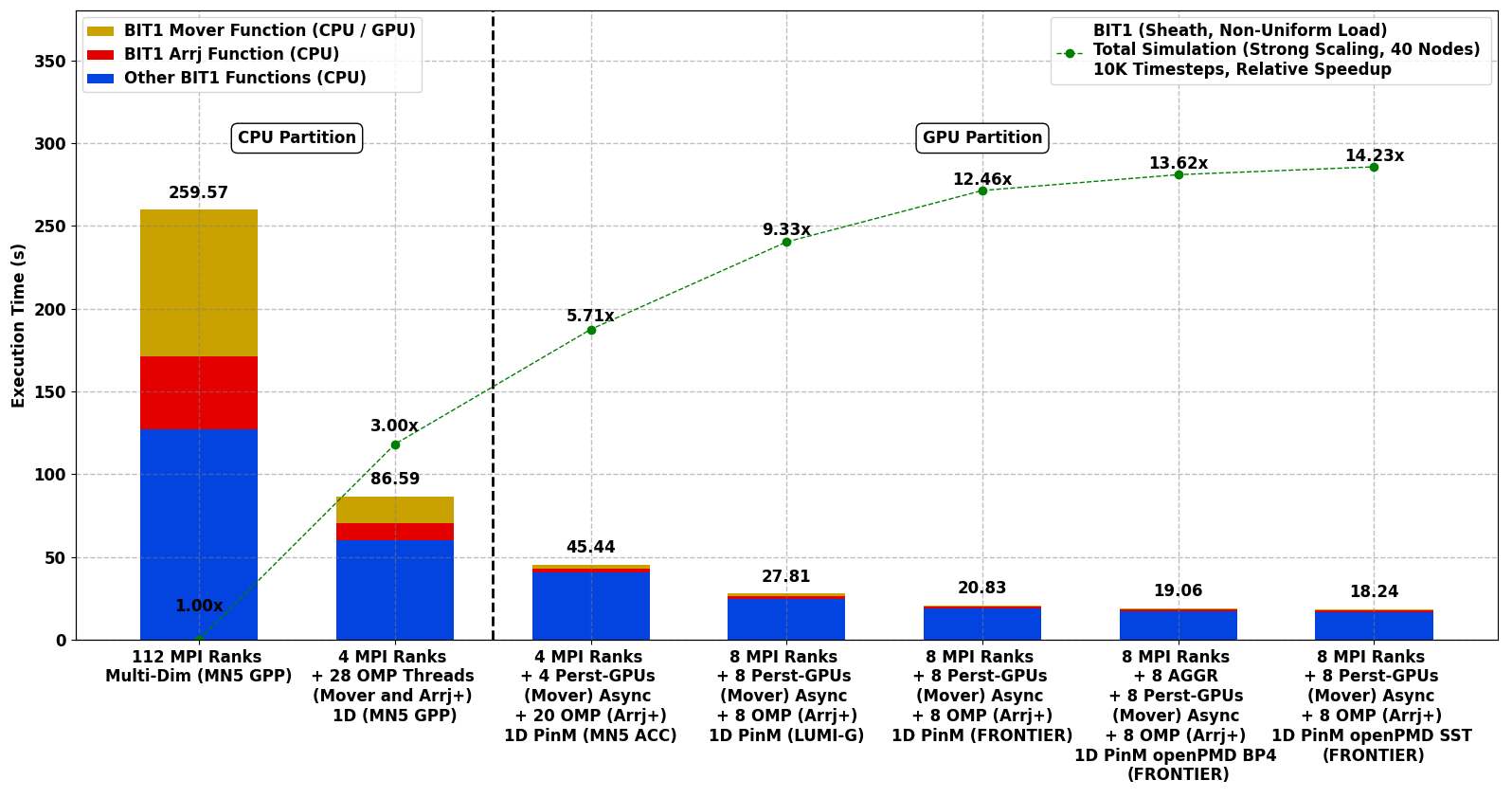}
        \caption{Hybrid BIT1 (Sheath, Non-Uniform Load) Total Simulation on 40 Nodes - Strong Scaling on MN5 GPP / ACC, LUMI-G, and Frontier for 40K timesteps.} 
        \label{BIT1_Sheath_Non_Uniform_Load_Strong_Scaling_40_Nodes_Execution_Time_UM_and_PinM_CPU_and_GPU}
    \end{center}
    \vspace{-0.6cm} 
\end{figure}

As shown in Fig.~\ref{BIT1_Sheath_Non_Uniform_Load_Strong_Scaling_40_Nodes_Execution_Time_UM_and_PinM_CPU_and_GPU}, the original BIT1 version on MN5 GPP (CPU partition), using 112 MPI ranks with a 3D data layout, executes for a total simulation time of $\approx$260~s, with the \texttt{mover} and \texttt{arrj} functions contributing $\approx$88~s and $\approx$44~s, respectively. Transitioning to a hybrid version with 4 MPI ranks and 28 OpenMP threads using a 1D layout reduces the total simulation time to $\approx$87~s (3.00$\times$ speedup), lowering the \texttt{mover} and \texttt{arrj} times to $\approx$16~s and $\approx$11~s. Moving to GPU-enabled executions with a 1D layout and PinM, the hybrid BIT1 version on MN5 GPP (GPU partition), using 4 MPI ranks and 4 persistent GPUs, with the \texttt{mover} using OpenMP target tasks asynchronously and OpenMP parallel \texttt{arrj}, further reduces the total time to $\approx$45~s (5.71$\times$ speedup), with \texttt{mover} and \texttt{arrj} at $\approx$3~s and $\approx$2~s. Scaling to 8 MPI and 8 persistent GPUs on LUMI-G improves performance further, achieving a total time of $\approx$28~s (9.33$\times$ speedup), with $\texttt{mover} \approx$2~s and $\texttt{arrj} \approx$1~s. On FRONTIER, the hybrid BIT1 GPU version with 8 MPI and 8 persistent GPUs and OpenMP thread parallelism reduces the total simulation time to $\approx$21~s (12.46$\times$ speedup), with $\texttt{mover} \approx$0.65~s and $\texttt{arrj} \approx$1~s. Enabling openPMD with the ADIOS2 BP4 backend introduces a slight improvement to a total time of $\approx$19~s (13.62$\times$ speedup), while openPMD SST further reduces it to $\approx$18~s (14.23$\times$ speedup), with the \texttt{mover} reaching $\approx$0.63~s and \texttt{arrj} $\approx$1~s, representing the fastest version across all systems.

\subsection{Accelerating \&  Scaling Hybrid BIT1 up to 800 GPUs}
Due to the considerable time required for non-uniform plasma conditions to reach equilibrium between particle sources and sinks~\cite{tskhakaya2010pic,vass2022revisiting}, a uniform load case is employed to enable immediate testing and analysis under controlled, ideal conditions. We evaluate a portable, multi-GPU hybrid MPI\allowbreak+OpenMP asynchronous version of BIT1 with scalable PLB and C/R mechanisms for high-performance resilience under uniform load conditions by performing a 2K timestep C/R sheath simulation, initialized from a prior 10K timestep run, on Frontier using up to 100 nodes (up to 800 GPUs), with parameters listed in Table~\ref{tab:bit1_input_parameters}. 

\begin{figure}[!ht]
    \vspace{-0.4cm} 
    \begin{center}
        \includegraphics[width=\linewidth]{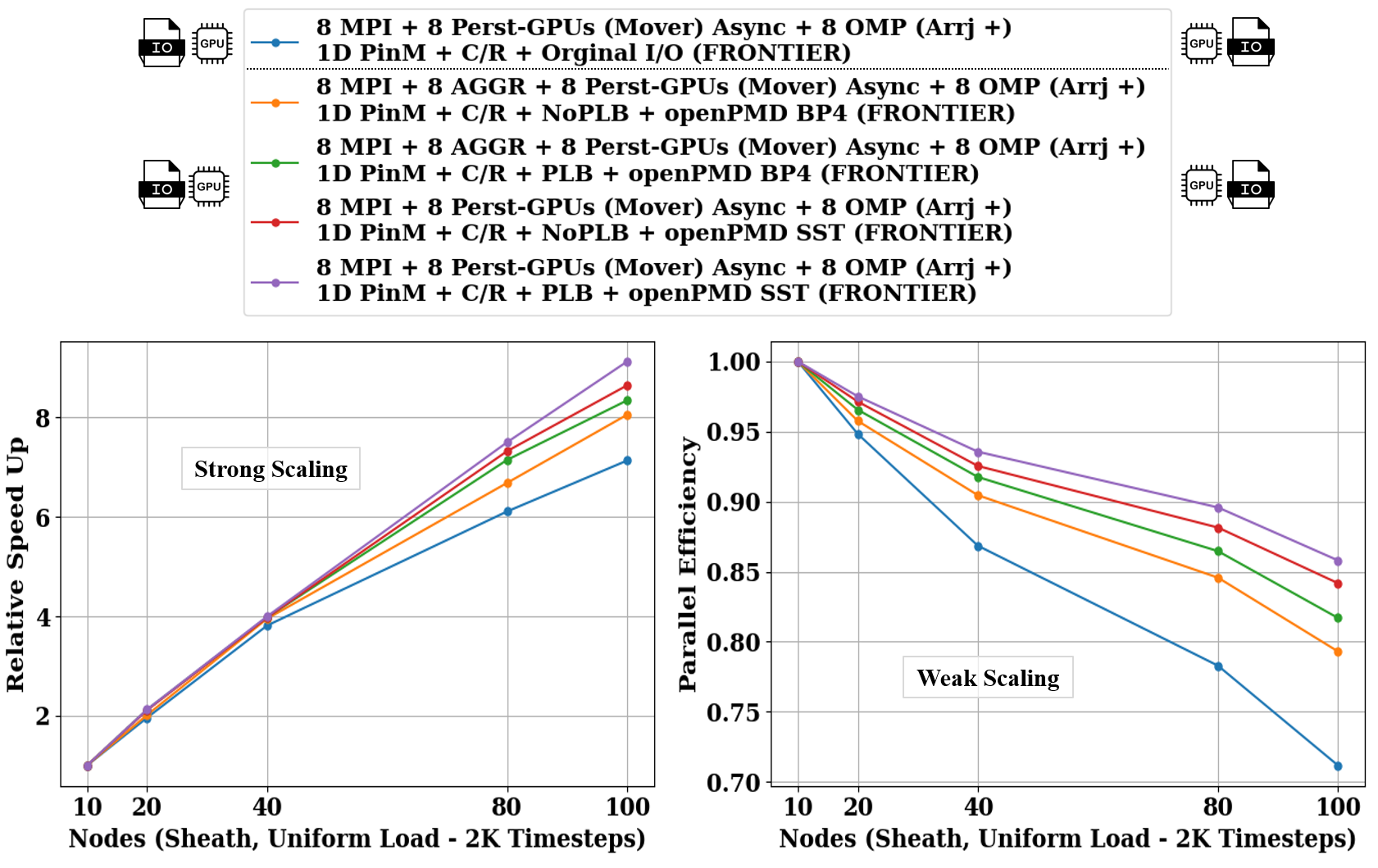}
        \caption{Hybrid BIT1 (Sheath, Uniform Load) C/R with PLB Total Simulation (Relative) Speed Up (left) and PE (Right) - Strong and Weak Scaling up to 100 Nodes (up to 800 GPUs) on Frontier for 2K timesteps.} 
        \label{BIT1_Sheath_Uniform_Load_Strong_and_Weak_Scaling_Frontier_Checkpoint_Restart_Review}
    \end{center}
    \vspace{-0.6cm} 
\end{figure}

As seen in Fig.~\ref{BIT1_Sheath_Uniform_Load_Strong_and_Weak_Scaling_Frontier_Checkpoint_Restart_Review} and strong scaling tests on Frontier, the original hybrid BIT1 (serial I/O) GPU version with C/R achieves a speedup of 7.14× when scaling from 10 to 100 nodes. Enabling openPMD, which introduces parallel I/O, with the ADIOS2 BP4 backend without PLB improves scalability to 8.06×, while enabling PLB further enhances performance to 8.35×. Using the ADIOS2 SST backend without PLB provides additional gains, reaching 8.65× speedup, and the combination of ADIOS2 SST with PLB delivers the best performance, achieving a 9.13× speedup. These results show that integrating openPMD with scalable C/R mechanisms improves strong scaling under heavy I/O conditions, with SST and PLB providing the best performance and scalability for large runs. 

For weak scaling, the original hybrid BIT1 GPU version with C/R exhibits a parallel efficiency (PE) of 70.8\% at 100 nodes, while the openPMD GPU versions show steadily improved weak scaling, where the BP4 backend (without PLB) sustains 76.1\% PE, enabling PLB increases this to 80.0\% PE, the SST backend (without PLB) further improves performance to 83.1\% PE, and the combination of SST with PLB achieves the best weak scaling performance with 88.0\% PE. Despite the increasing workload with frequent time dependent output and checkpointing, the openPMD (BP4 and SST) GPU versions consistently maintain higher parallel efficiency than the original hybrid BIT1 GPU version, demonstrating improved weak scaling behavior and greater resilience to I/O and C/R overheads at scale.

\subsection{Hybrid BIT1 C/R + PLB In-Situ Analysis \& Visualization}
A key aspect of this work is enabling real-time checkpoint analysis and visualization during large-scale non-uniform sheath simulations without interrupting execution. This is achieved through integration of openPMD with the ADIOS2 BP4 and SST backends, enabling file-based and streaming in-situ workflows.

\begin{figure}[!ht]
\vspace{-0.4cm} 
    \begin{center}
        \includegraphics[width=\linewidth]{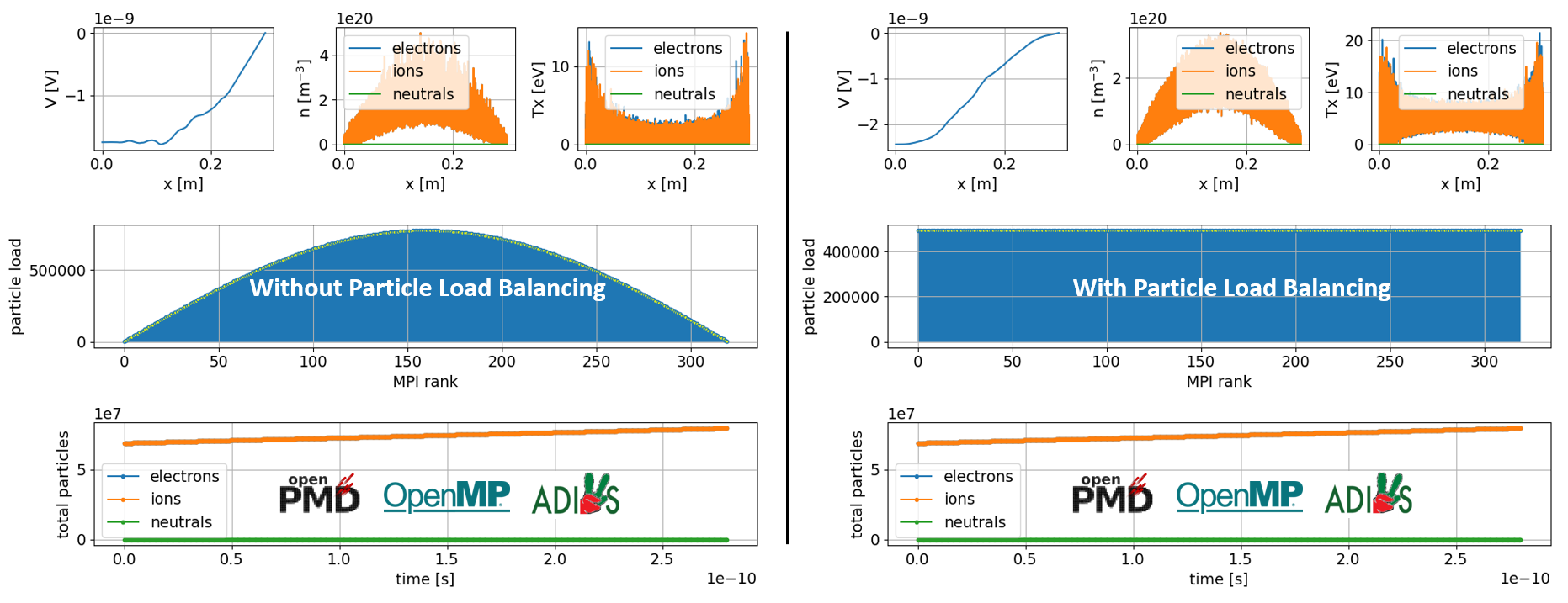}
        \caption{Performing real-time checkpoint data analysis and visualization using the hybrid BIT1 openPMD BP4/SST (sheath, non-uniform load) simulation on Frontier (with corresponding analysis and visualizations on Dardel GPU, LUMI-G and MN5 ACC), for up to 10K timesteps with 5 C/R on 40 Nodes (320 GPUs), without (left) and with (right) PLB, highlighting a reduced set of plasma profiles, the particle load per MPI rank, and the time evolution of the total number of particles for each species.} 
        \label{BIT1_Sheath_Non_Uniform_Load_Plot-5x_10K_Timesteps-Final_Checkpoint_Restart}
    \end{center}
    \vspace{-0.6cm} 
\end{figure}

Fig.~\ref{BIT1_Sheath_Non_Uniform_Load_Plot-5x_10K_Timesteps-Final_Checkpoint_Restart} shows simultaneous monitoring of the hybrid BIT1 openPMD BP4/SST (sheath, non-uniform load) simulation with and without PLB on Frontier. On the left, without PLB, each MPI rank processes a fixed block of cells, which can lead to uneven particle distributions, whereas on the right PLB dynamically redistributes particles across ranks and updates the cell ranges accordingly prior to parameterization and memory allocation. 



\section{Discussion \& Conclusion}
This work combines portability, resilience, and scalable I/O, which are essential for efficient large-scale PIC MC simulations on modern heterogeneous HPC systems. The hybrid MPI+OpenMP multi-GPU BIT1 achieves significant performance gains through a 1D data layout, pinned memory, persistent GPU allocation, and asynchronous execution, reducing overhead across Nvidia and AMD architectures. Integrating openPMD with ADIOS2 (BP4 and SST) improves scalability under heavy I/O and C/R workloads and enables improved file-based I/O, streaming I/O, and dynamic workload redistribution at scale. 

Non-uniform load sheath simulations confirm that PLB is critical for mitigating load imbalance caused by uneven particle distributions, improving runtime stability and resource utilization, while in-situ analysis and visualization enable real-time monitoring of plasma behavior and workload distribution without interrupting execution, reducing reliance on costly post-processing.

Future work will extend hybrid BIT1 simulations to Intel GPU platforms at exascale, targeting \texttt{Aurora} and Europe’s first exascale system, \texttt{JUPITER Booster}, to evaluate portability and resilience across Nvidia, AMD, and Intel architectures, and to identify any remaining architecture specific limitations.


\vspace{2mm} 
\noindent \footnotesize{\textbf{Acknowledgments.} Funded by the European Union. This work has received funding from the European High Performance Computing Joint Undertaking (JU) and Sweden, Finland, Germany, Greece, France, Slovenia, Spain, and Czech Republic under grant agreement No 101093261 (Plasma-PEPSC). The computations/data handling were/was enabled by resources provided by the National Academic Infrastructure for Supercomputing in Sweden (NAISS), partially funded by the Swedish Research Council through grant agreement no. 2022-06725. This research used resources of the Oak Ridge Leadership Computing Facility at the Oak Ridge National Laboratory, which is supported by the Advanced Scientific Computing Research programs in the Office of Science of the U.S. Department of Energy under Contract No. DE-AC05-00OR22725





\bibliographystyle{splncs04}
\bibliography{bit1paper.bib}





\end{document}